\documentclass[runningheads]{cl2emult}

\usepackage{makeidx}  
\usepackage{graphicx} 
\usepackage{subeqnar} 
\usepackage{multicol} 
\usepackage{cropmark} 
\usepackage{eso}      
\usepackage{psfig}
\makeindex            

\begin{document}
\title*{Chemical Yields from the First Stars}
\toctitle{Chemical Yields from the First Stars}
%
%
%
\titlerunning{Chemical Yields from the First Stars}
%
\author{Laura Portinari}
\authorrunning{L.~Portinari}
%
%
\institute{Dipartimento di Astronomia, Universit\`a di Padova\\
Vicolo dell'Osservatorio 5, I-35122 Padova (PD), Italy}

\maketitle              


\begin{abstract}
We examine the dependence of stellar yields on the metallicity $Z$ 
of the stellar population. 
This effect may be important for
the very first 
chemical enrichment from Population III stars, at very low $Z$. 
In the range of massive stars, mass loss 
rates varying with $Z$ have remarkable effects. 
We also estimate chemical yields from Very Massive Objects 
(from 120 to 1000~$M_{\odot}$), which might have formed more easily 
in the very low-$Z$ environment of the first stellar generations.
\end{abstract}


\section{Introduction}

Stellar model calculations show that the detailed structure and evolution 
of a star of given mass depends on its chemical composition;
so we expect stellar yields as well to be influenced by metallicity.
A 
homogeneous set of metallicity dependent yields,
covering the whole stellar mass range, has been derived by the Padova group. 
Detailed results can be found in~\cite{PCB98} (hereinafter PCB98) 
for massive and very massive stars and in~\cite{Marigoetal96,Marigoetal98} 
for low and intermediate mass stars.  
%
In this paper we will discuss the yields from massive and very massive
stars; full details can be found in PCB98.


\section{Yields from Massive Stars ($M=6-120 \, M_{\odot}$)}

In the range of massive stars, quiescent mass loss 
has
a strong influence on the yields, because (1) ejecta are directly released
through the stellar wind,
and (2) 
mass loss affects the final total and core mass and thus also, indirectly,
the final supernova (SN). The efficiency of radiation pressure driven wind
scales with metallicity ({\mbox{$\dot{M} \propto Z^{0.5}$}}
\cite{Kudritzkietal89}). So, through the mass loss rate metallicity
affects the mass and composition of the layers peeled off in the wind,
as well as the final stellar mass $M_{fin}$ and CO-core mass $M_{CO}$
(Fig.~1). Notice how the most massive stars ($M \geq 40 \, M_{\odot}$) in the
high metallicity sets end up with similar, low final masses
($4-6 \, M_{\odot}$). These stars go through a WR stage, where mass loss is
efficient and strongly mass-dependent ({\mbox{$\dot{M} \propto M^{2.5}$}}
\cite{Langer89}): their mass decreases rapidly while $\dot{M}$
also decreases correspondingly, until they all reach
very similar final masses. The switch to the WR stage at high masses
produces a peak in $M_{fin}$ and $M_{CO}$, which is less and less prominent,
and corresponds to lower and lower initial masses, the higher the metallicity
(Fig.~1). For the lowest $Z$, the peak mass falls in 
the range of 
very massive stars, where this behaviour is qualitatively extended (Sect.~3).

Basing on the Padova tracks \cite{Bressanetal93,Fagottoetal94a,Fagottoetal94b}
PCB98 calculated stellar yields of massive stars ($6-120 \, M_{\odot}$) with 
mass loss for 5 sets of different metallicity, from $Z=0.0004$ to $Z=0.05$.
%
%
The contribution to stellar yields from hydrostatic 
stages beyond He-burning 
and from explosive nucleosynthesis of {\it iron-core collapse} SN was estimated
by matching our stellar pre-SN models with the SN yields by~\cite{WW95}, 
on the base of the respective CO-core mass $M_{CO}$.

\begin{figure}[t]
\hspace{-1.5truecm}
{\centering \leavevmode
\psfig{file=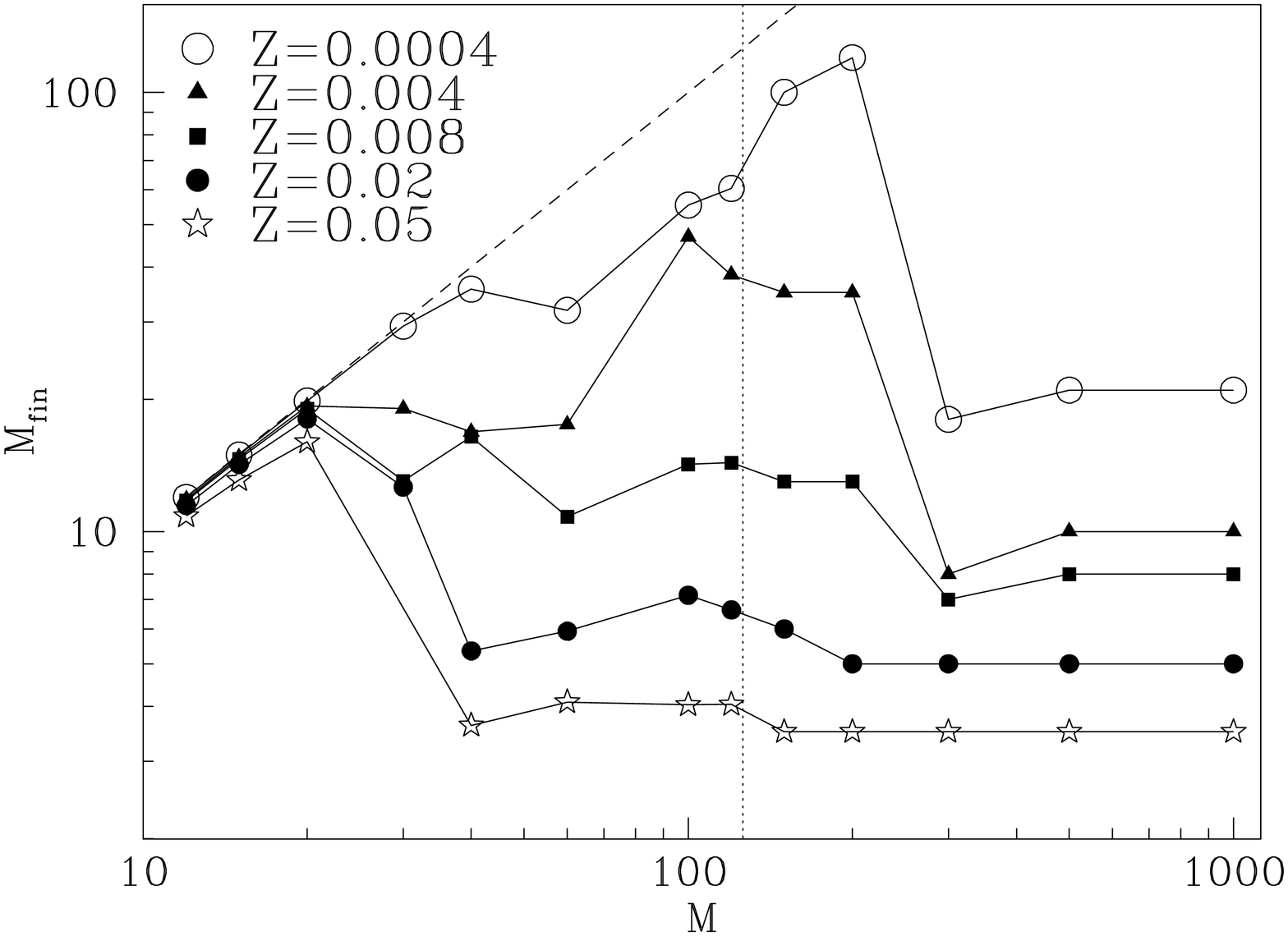,angle=-90,width=.65\textwidth} \hfil
\psfig{file=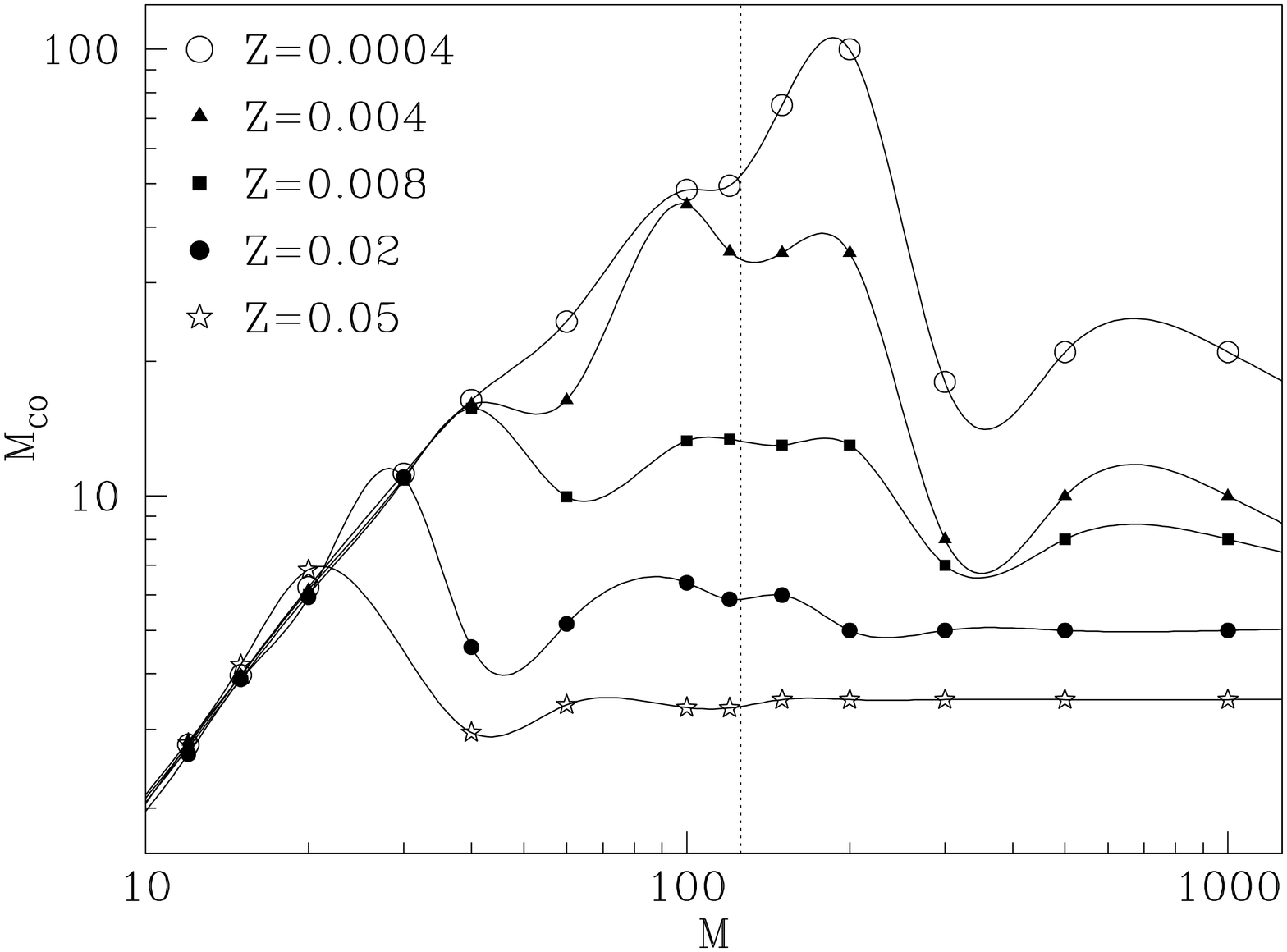,angle=-90,width=.65\textwidth}}
\caption{{\it Left panel}: Final mass vs.\ initial stellar mass for five 
metallicities. The dashed line corresponds to constant
mass evolution. The vertical dotted line separates massive stars with
detailed stellar tracks (Sect.~2) from VMOs (qualitative calculations,
Sect.~3). {\it Right panel}: Core mass vs.\ initial mass.}
\end{figure}

At the high mass end, {\it pair creation} (PC) SN\ae\ are expected. 
Following~\cite{Woosley86}, depending on the detailed mass the outome may be: 
partial 
explosion with ejection of some layers and the rest falling into a black hole
($M_{He}=35-60 \, M_{\odot}$), total thermonuclear explosion 
($M_{He}=60-110 \, M_{\odot}$), complete collapse into a black hole 
($M_{He}>110 \, M_{\odot}$); see also~\cite{Heger99}.
In our stellar models, mass loss inhibits the growth of so large cores,
and {\mbox{PC SN\ae}} are confined to the most massive stars in the 
lowest metallicity range ($M \geq 100 \, M_{\odot}$ and $Z \leq 0.004$). 
The adopted SN yields for these PC cases are from~\cite{Woosley86}.

The fractionary stellar yields of massive (and very massive, see Sect.~3) 
stars 
are shown in Fig.~2. At low metallicities,
remnant masses can be very large due to low mass-loss rates: massive cores 
are built, leaving large remnants; and 
when most of the core mass falls into a black hole, little oxygen is released.
For the largest masses ($M \geq 100 \, M_{\odot}$), PC SN\ae\ occur
and the oxygen yields increase again, while remnant masses decrease. 

For metallicities $Z \geq 0.008$ the effects of mass loss become apparent:
for $M \geq 30 \, M_{\odot}$, remnants 
are much smaller than in the 
low-$Z$ sets, while helium and carbon yields increase because of the
contribution of the wind.
 
For heavy elements (Si, S, Ca, Fe) the bulk of contribution generally comes 
from stars in the range $10-30 \, M_{\odot}$.


\section{Yields from Very Massive Stars ($M=120-1000 \, M_{\odot}$)}

A primeval Population III of very massive objects (VMOs)
was invoked in past years as a possible 
solution for the G-dwarf problem and for the non-zero metallicity of 
Population II stars, to form black holes which could account for Dark Matter 
and AGNs, to explain the reionization of the Universe, to produce primordial
helium; for a review see~\cite{Chiosi99}.
Various stellar models for VMOs were therefore developed in the '80s 
\cite{Bondetal84,Eleidetal83,Oberetal83,Woosley&Weaver82,Klapp83,Klapp84}.

VMOs might have formed more easily at the low metallicity of the early epochs,
when the Initial Mass Function (IMF) was probably more top--heavy;
e.g.\ \cite{Brommetal99tc,Brommetal99,Chiosi99}. But the interest for VMOs
is not limited to extremely low-$Z$ environments: the Pistol star with
$M = 200-250 \, M_{\odot}$ has been discovered in the Galactic Centre,
 where $Z \geq Z_{\odot}$ \cite{Figeretal98}; and other such objects
are known within the Local Group \cite{Kudritzki97}.
Therefore, for any $Z$ it is of interest to extend the grid of stellar yields
beyond $M=120 \, M_{\odot}$, which we did in qualitative terms (PCB98).

Basing on the above mentioned papers on VMOs 
and on considerations of continuity with our stellar tracks for massive stars,
we 
expect the structure and evolution of VMOs to be as follows. 
The H--burning lifetime of VMOs is $2-3 \, Myr$, and H--burning takes place 
in the inner 50\% of their initial mass. During H--burning VMOs undergo
pulsational instability with violent mass loss, maybe
as high as $10^{-3} \, M_{\odot}/yr$, independent on $Z$.
Their mass therefore falls rapidly (in $10^5 - 10^6 \, yrs$)
below $120 \, M_{\odot}$; from then on, they will end the phase of
paroxysmal mass loss, enter the normal regime of radiation pressure driven 
wind and follow roughly the fate of a star of $100-120 \, M_{\odot}$ for the 
corresponding $Z$. At very high masses, the (large!) H--burning core may be 
at some point revealed on the surface: the star then becomes a WR of 
very large mass and thus very large $\dot{M}$ (see Sect.~2), and decreases
to a rather small final mass (Fig.~1).
The core mass $M_{CO}$ eventually drives the final SN explosion,
as assumed for massive stars. In most cases the outcome is an iron--core 
collapse SN; only for very low $Z$ a few PC SN\ae\ are found.

This gross scenario gives us an analytical estimate of the yields of VMOs;
for details see PCB98. The grid of stellar yields can thus be extended 
up to $1000 \, M_{\odot}$ for the 5 metallicities; the results are shown
in Fig.~2.
For the lowest metallicity, $Z=0.0004$, stars with $150-200 \, M_{\odot}$ 
enter the regime of PC SN\ae\ with complete thermonuclear disruption
($M_{CO}=60-110 \, M_{\odot}$) and large release of oxygen and heavy elements.
If the IMF at low metallicities is skewed toward very high 
masses, these objects might contribute substantially to the
very early chemical enrichment of galaxies (as pointed out also by 
\cite{Langer&Woosley96}). 
At even larger masses, mass loss during 
the large WR stage reduces the final stellar and core mass to low values, 
so that the ejecta mainly consist
of helium lost through the wind, with little production of metals. 

This latter
behaviour holds for all VMOs at higher $Z$, due to stronger mass loss
in the radiation pressure wind phase (after they fall below 
$120 \, M_{\odot}$).

\begin{figure}[ht]
{\centering \leavevmode
\psfig{file=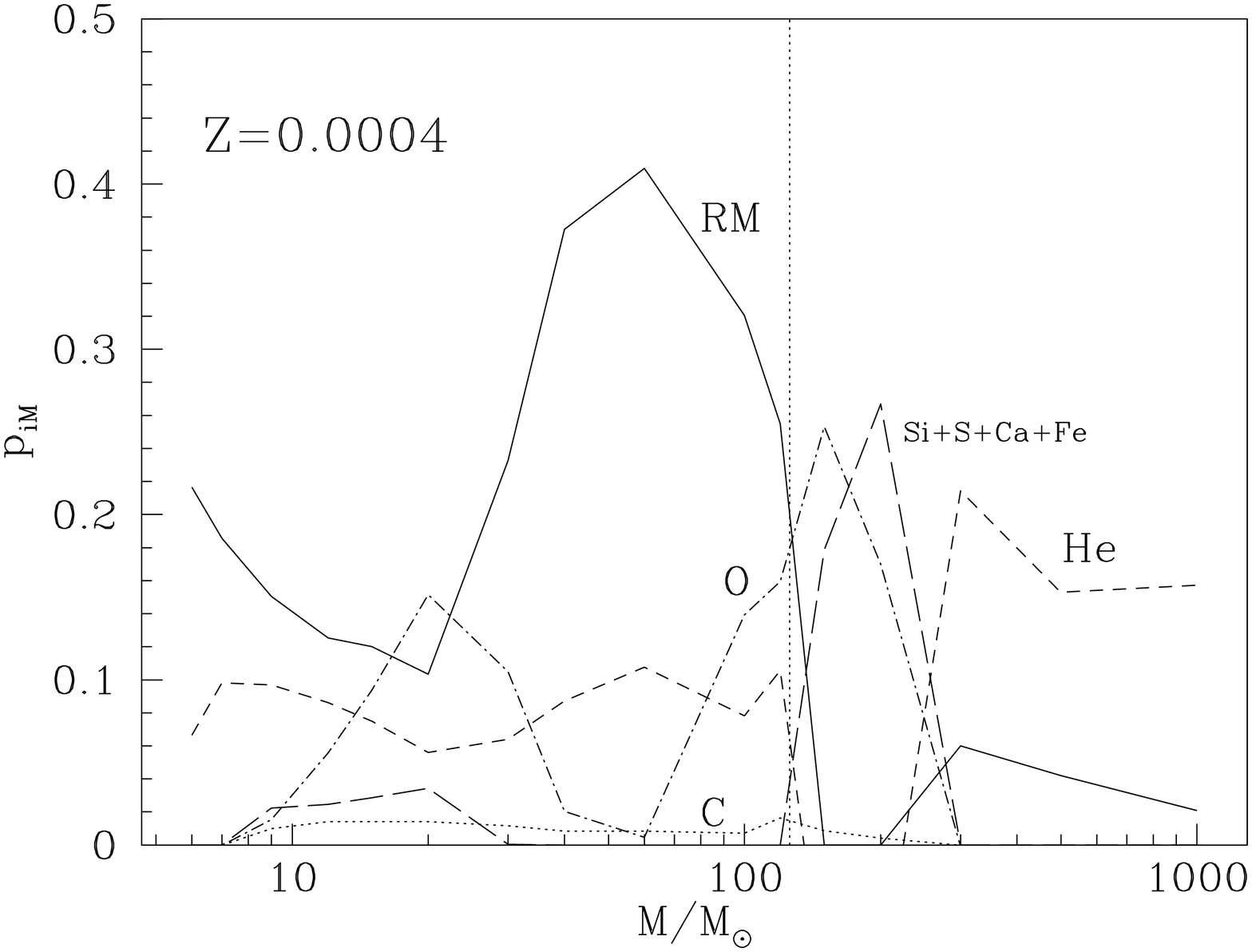,angle=-90,width=\textwidth}}

\hspace{-1.5truecm}
{\centering \leavevmode
\psfig{file=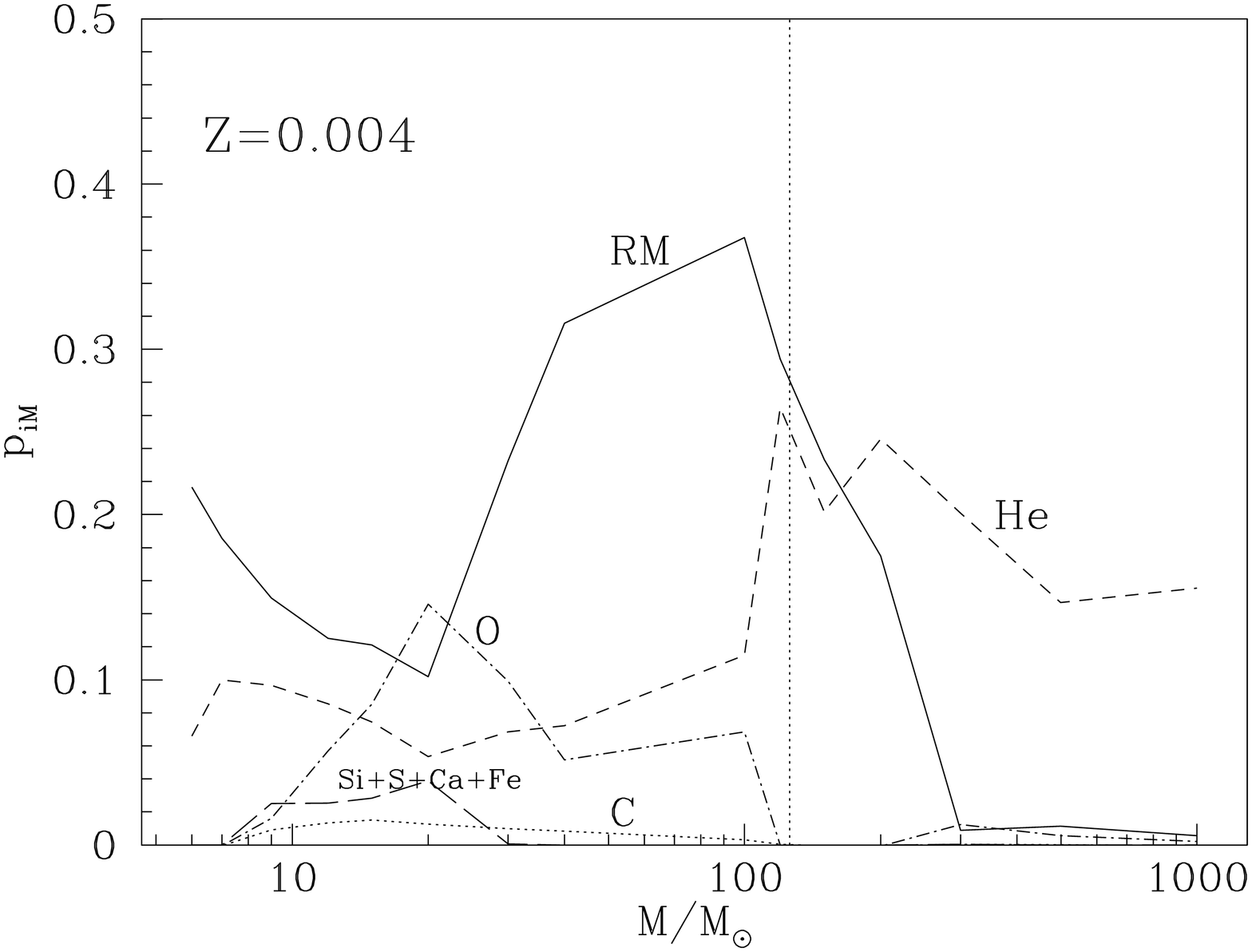,angle=-90,width=.65\textwidth} \hfil
\psfig{file=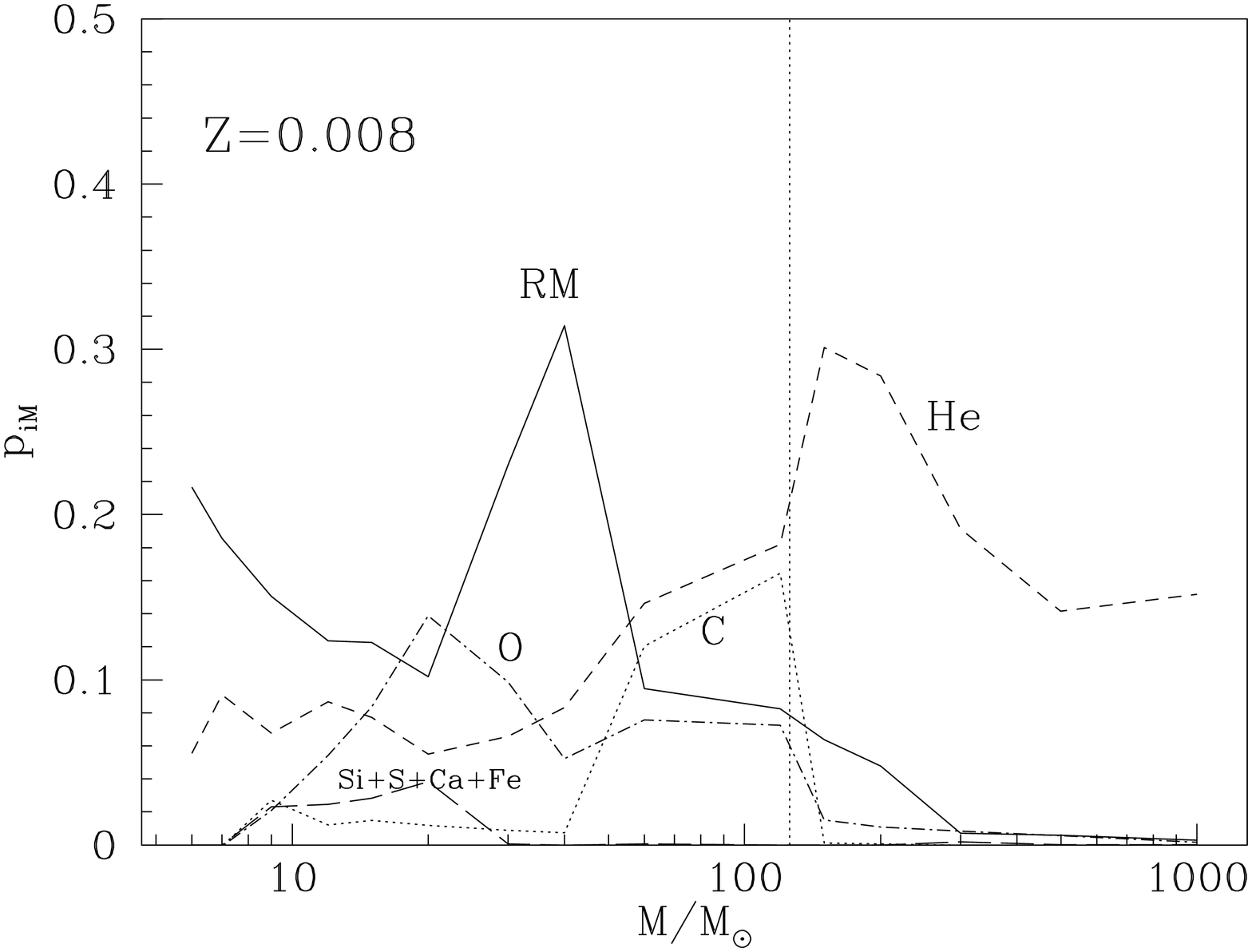,angle=-90,width=.65\textwidth}}

\hspace{-1.5truecm}
{\centering \leavevmode
\psfig{file=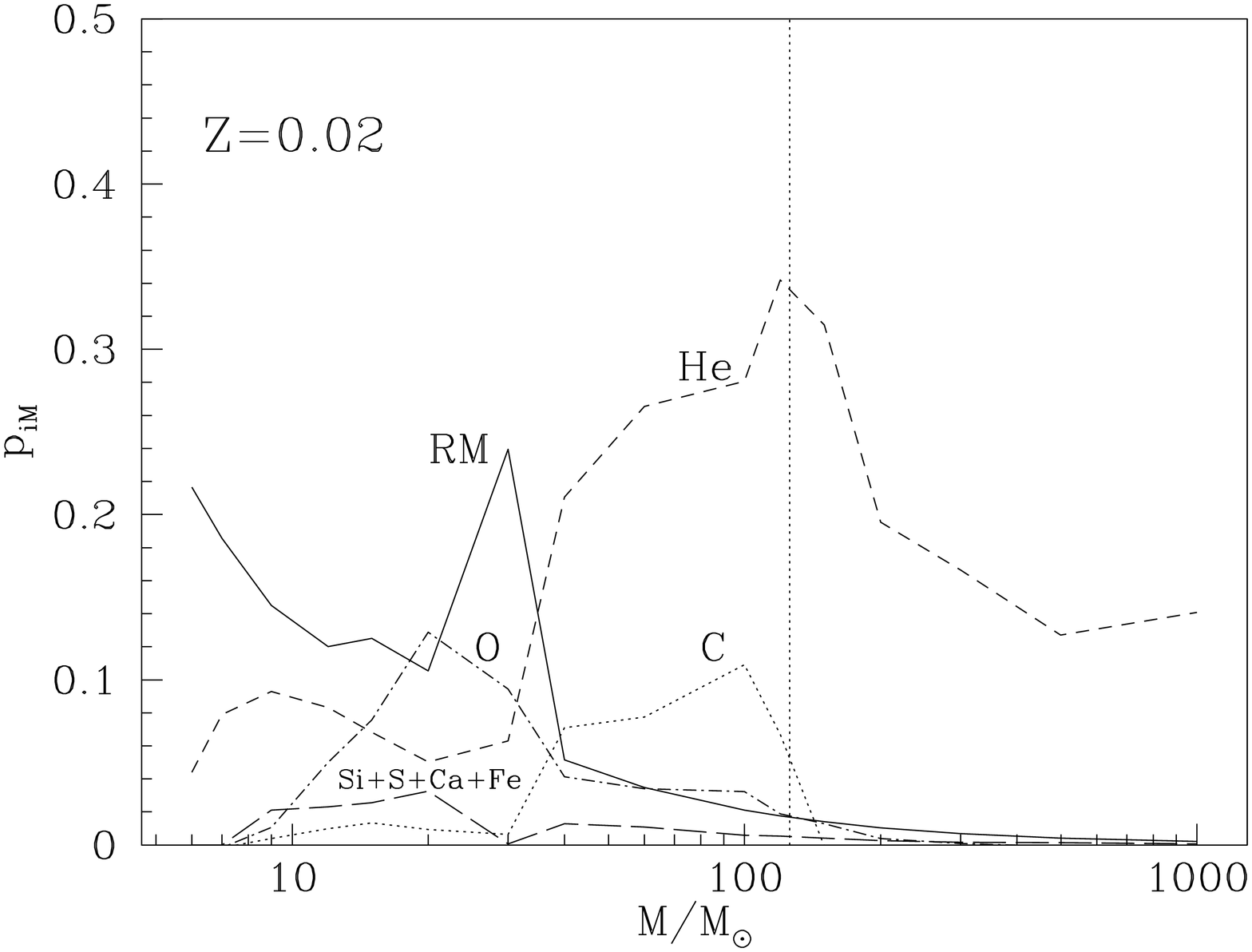,angle=-90,width=.65\textwidth} \hfil
\psfig{file=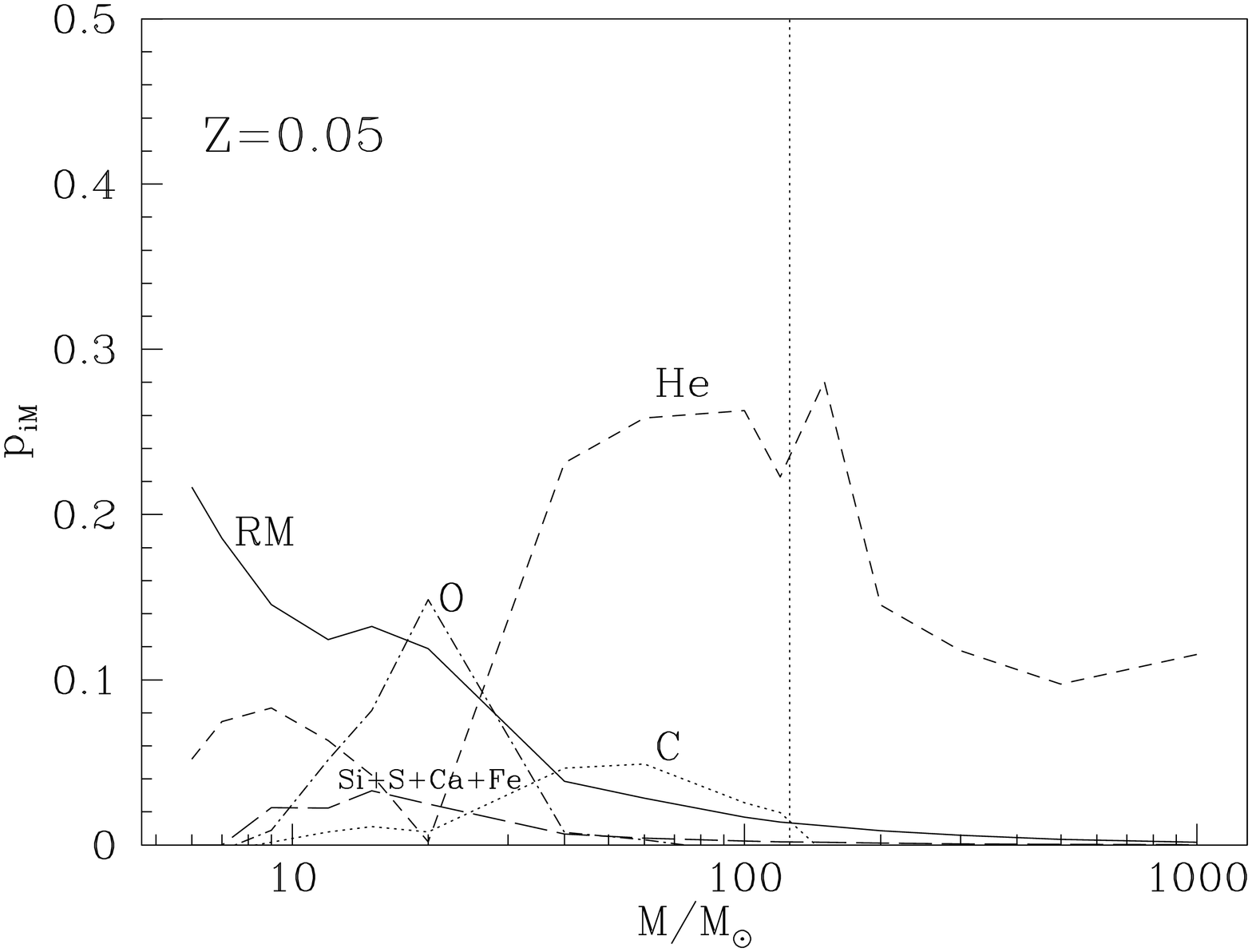,angle=-90,width=.65\textwidth}}
\caption{Fractional remnant mass RM=$M_r/M$ and fractional stellar yields
of a few elements for massive stars and VMOs of different metallicities.
Solid line: RM; dashed line: helium yields; dotted line: carbon yields;
dash-dotted line: oxygen yields; long-dashed line: heavy elements yields.
The vertical dotted line separates massive stars with
detailed stellar tracks (Sect.~2) from VMOs (qualitative calculations,
Sect.~3).}
\end{figure}

\clearpage

Of course, the gross behaviour of VMOs is expected to depend on the 
efficiency of mass loss, which is basically unknown for the regime of 
pulsational instability typical of VMOs, and for very low $Z$ in general
\cite{Chiosi99}. What if the assumed
mass loss rate in the violent phase is decreased, say, from $10^{-3}$ 
to $10^{-4} \, M_{\odot}/yr$?
In most cases, such a mass loss rate is still fast enough to reduce
the stellar mass below $120 \, M_{\odot}$ in a short time, and then the 
overall evolution will remain substantially the same. Only for VMOs of 
$500-1000 \, M_{\odot}$ the scenario
will change: losing only $200-300 \, M_{\odot}$ during their lifetime,
they never become WR stars and result in a final core mass of $\sim 250$
and $500 \, M_{\odot}$ respectively, ending up in a black hole collapse.

Calculations on stellar evolution and yields down to $Z$=0 are under 
way~\cite{Marigoetal99}.

\subsubsection{Acknlowledgements}
I acknowledge financial support from MPA during this conference.
This study is financed by the Italian MURST, contract n.~9802192401.

\clearpage
\addcontentsline{toc}{section}{Index}
\flushbottom
\printindex

\end{document}